\documentclass[journal=jacsat]{achemso}
\setkeys{acs}{articletitle = true}

\usepackage[dvipsnames]{xcolor}
\usepackage{color}
\usepackage[version=3]{mhchem} 

\usepackage{times}
\usepackage{amsmath}
\usepackage{amssymb}
\usepackage{bm}
\usepackage{here}
\usepackage{graphicx}
\usepackage{framed}
\definecolor{shadecolor}{rgb}{0.9,0.9,0.9}
\usepackage[utf8x]{inputenc}

\author{Nisha Geng}
\affiliation[State University of New York at Buffalo]
{Department of Chemistry, State University of New York at Buffalo, Buffalo, NY 14260-3000, USA}
\author{Katerina P. Hilleke}
\affiliation[State University of New York at Buffalo]
{Department of Chemistry, State University of New York at Buffalo, Buffalo, NY 14260-3000, USA}
\author{Li Zhu}
\affiliation[Rutgers University, Newark]
{Department of Physics, Rutgers University, Newark, NJ 07102, USA}
\author{Xiaoyu Wang}
\affiliation[State University of New York at Buffalo]
{Department of Chemistry, State University of New York at Buffalo, Buffalo, NY 14260-3000, USA}
\author{Timothy A. Strobel}
\affiliation[Carnegie Institution of Washington]
{Earth and Planets Laboratory, Carnegie Institution of Washington, Washington, DC 20011, USA}
\author{Eva Zurek}
\affiliation[State University of New York at Buffalo]
{Department of Chemistry, State University of New York at Buffalo, Buffalo, NY 14260-3000, USA}
\email{ezurek@buffalo.edu}

\title{Conventional high-temperature superconductivity in metallic, covalently bonded, binary-guest C-B clathrates }

\begin{document}

\clearpage
\newpage
\begin{abstract}
Inspired by the synthesis of $X$B$_3$C$_3$ ($X$=~Sr, La) compounds in the bipartite sodalite clathrate structure, density functional theory (DFT) calculations are performed on members of this family containing up to two different metal atoms. A DFT-chemical pressure analysis on systems with $X$=~Mg, Ca, Sr, Ba reveals that the size of the metal cation, which can be tuned to stabilize the B-C framework, is key for their ambient-pressure dynamic stability. High-throughput density functional theory calculations on 105 $Pm\bar{3}$ symmetry $XY$B$_6$C$_6$ binary-guest compounds (where $X, Y$ are electropositive metal atoms) find 22 that are dynamically stable at 1~atmosphere, expanding the number of potentially synthesizable phases by 19 (18 metals and 1 insulator). The density of states at the Fermi level and superconducting critical temperature, $T_c$, can be tuned by changing the average oxidation state of the metal atoms, with $T_c$ being highest for an average valence of +1.5.  KPbB$_6$C$_6$, with an ambient-pressure Eliashberg $T_c$ of 88~K, is predicted to possess the highest-$T_c$ among the studied $Pm\bar{3}n$ $X$B$_3$C$_3$ or $Pm\bar{3}$ $XY$B$_6$C$_6$ phases, and calculations suggest it may be synthesized using high-pressure high-temperature techniques then quenched to ambient conditions. 
\end{abstract}

\newpage
\section{Introduction}

The advent of the high-pressure superconducting hydrides has renewed interest in conventional superconductors, demonstrating that their critical temperatures ($T_c$s) may approach room temperature~\cite{Zurek:2021k}. The discovery of many of these compounds was theory driven, highlighting that first principles-based methods -- consisting of crystal structure prediction (CSP) searches and electron-phonon coupling calculations  -- can identify promising superconducting materials for future syntheses~\cite{Zurek:2018m,Zurek:2020k,Flores-Livas:2020}. One of the structure types that has emerged to be conducive for superconductivity, initially pinpointed theoretically within hydrides containing an electropositive metal \cite{Peng:2017a,Ma:2012f}, is an $Im\bar{3}m$ symmetry $X$H$_6$  superhydride. This structure type is based on a bcc packing of face-sharing $X$@H$_{24}$ truncated octahedra with six square and eight hexagonal faces, in which the hydrogenic lattice, isostructural with the sodalite-type clathrate, is stuffed with alkaline-earth or rare-earth metal atoms. Theory has identified many stable compounds possessing this motif -- and recently a number of them have been  synthesized under high pressure and their $T_c$s have been measured  (e.g., CaH$_6$ [215 K at 172 GPa\cite{Ma:CaH6}, 210 K at 160 GPa\cite{Li:CaH6}], YH$_6$ [220 K at 183 GPa\cite{Kong:2021a}, 224 K at 166 GPa\cite{Troyan:2021a}], and (La,Y)H$_6$, with a transition circa 237~K at 183~GPa tentatively attributed~\cite{Semenok:2021}).

\begin{figure*}[ht!]
\begin{center}
\includegraphics[width=12cm]{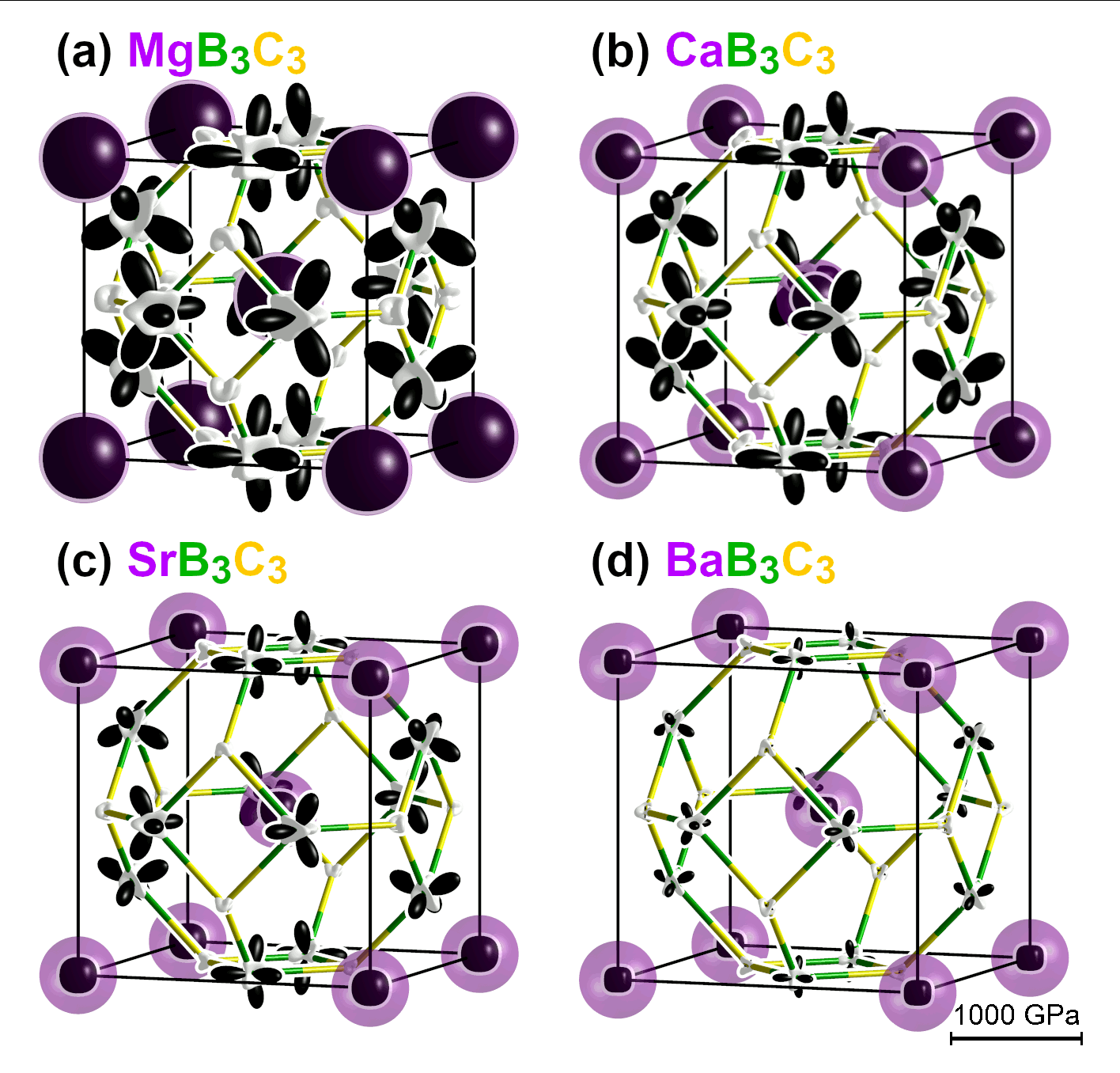}
\end{center}
\caption{Chemical pressure (CP) schemes of simple B-C clathrates in the bipartite sodalite $Pm\bar{3}n$ structure at ambient pressure: (a) MgB$_3$C$_3$, (b) CaB$_3$C$_3$, (c) SrB$_3$C$_3$ and (d)~BaB$_3$C$_3$. Chemical pressures are represented by atom-centered spherical harmonic functions in which the magnitude of the CP in a particular direction is represented by the size of the lobes and their color (white = positive and black = negative). A scalebar representing the magnitude of the CP lobes is included. The large negative pressures on the Mg atom demonstrate a poor fit inside the surrounding borocarbide cage, which improves (with a concomitant decrease in the negative CPs) as the size of the metal atoms increase from Ca to Sr to Ba. Metal atoms are denoted in purple, and B/C atoms emanate green/yellow stick bonds.}
\label{fig:MB3C3_CP}
\end{figure*}

Achieving high-temperature superconductivity, even room-temperature superconductivity, is therefore no longer the `holy grail'. But, because none of the predicted or synthesized high-$T_c$ superhydrides are stable (or even metastably recoverable) at ambient pressure, the immediate challenge is to find light-element-based structural analogues or derivatives that could be metastable at 1~atm. One class of materials that are actively being considered\cite{Li:2020sr,zhu:arxiv,Strobel:2021la,Wang:2021,Di:2022,Zhang:2022,Gai:2022,Zhu:2020ferr} can be constructed from the $X$H$_6$ superhydride lattices by replacing the framework hydrogen atoms by carbon and boron atoms (some examples are illustrated in Fig.\ \ref{fig:MB3C3_CP}). In these light-element hexahydride analogues the clathrate lattice is held together by strong covalent B-C bonds, which are metallized via electron transfer from an encapsulated metal atom with the appropriate valence. Vibrations of the covalently bonded metallic lattice induces electron-phonon coupling, resulting in superconductivity. Indeed, efficient electron-phonon coupling has been predicted and observed in various sp$^2$/sp$^3$ covalent materials, with the most famous example being MgB$_2$ \cite{Blase:2009, Connetable:2003, Connetable:2004, Zipoli:2006, Breda:2000c}.

Pure carbon clathrates have yet to be synthesized, and geometric constraints place strict limitations on the atoms that could potentially be stuffed into their cages \cite{Nesper:1993,Cros:2009,Rey:2008,Alyssa:2002}. However, calculations have suggested that such materials can be stabilized by substituting some of the carbon atoms with boron and inserting small cations into the borocarbide framework \cite{Zeng:2015a}. Boron and carbon are the lightest elements that can form strong covalent bonds, and materials based on these elements are known to be good candidates for phonon mediated superconductivity at atmospheric conditions (e.g., MgB$_2$, $T_c=$~39~K \cite{mgb2}, boron doped diamond $T_c=$~4~K for a doping level of 2.5\%~\cite{Ekimov2004}, B-doped Q-carbon with a $T_c$ as high as 55~K~\cite{Bhaumik:2017}). $Pm\bar{3}n$ SrB$_3$C$_3$ is the first member of the family of borocarbide analogues of the clathrate hexahydrides to be predicted computationally via CSP techniques~\cite{Li:2020sr}. This compound, in the bipartite sodalite (Type-VII clathrate) structure, was computed to be thermodynamically stable between 50-200~GPa \cite{Li:2020sr} with a $T_c$ predicted to be as high as 43~K at 1~atm~\cite{zhu:arxiv}. Subsequently, SrB$_3$C$_3$ was synthesized at 57~GPa and quenched to ambient conditions in an inert atmosphere, and evidence for the superconducting transition was recently observed~\cite{Li:2020sr,zhu:arxiv}. Not long after, an isotypic lanthanum phase, LaB$_3$C$_3$, was synthesized at milder pressure and quenched to ambient conditions, where its HSE06 bandgap was computed to be 1.3~eV~\cite{Strobel:2021la}. 

The synthesis of SrB$_3$C$_3$ and LaB$_3$C$_3$ inspired numerous theoretical investigations of related materials. The superconducting mechanism in the hole-conductor $Pm\bar{3}n$ SrB$_3$C$_3$ was shown to result from the strong coupling between the $sp^3$  $\sigma$-bonding bands and boron-associated $E_g$ modes with a $T_c$ of 40~K, as estimated via solution of the Eliashberg equations.~\cite{Wang:2021} In fact, the $T_c$ calculated for all of the $Pm\bar{3}n$ $X$B$_3$C$_3$ alkaline earth analogues ($X=$~Ca, Sr, Ba) ranged from 40-50~K~\cite{Wang:2021,Di:2022, Zhang:2022,zhu:arxiv}. $Pm\bar{3}n$ ScB$_3$C$_3$ was dynamically unstable, however following the imaginary eigenvectors resulted in a non-centrosymmetric $Ama2$ structure with a spontaneous polarization that was large compared with other well known ferroelectric materials \cite{Zhu:2020ferr}. The superconducting properties of a few borocarbide clathrates stuffed with two different metal atoms have also been studied theoretically~\cite{Di:2022,Zhang:2022,Gai:2022}. In the silicon analogues, $Pm\bar{3}n$ RbB$_3$Si$_3$ was thermodynamically stable with respect to the elemental phases between 7-35~GPa, and it remained metastable at 1~atm with an estimated $T_c$ of 14~K \cite{Cui:2020}. CSP coupled with high-throughput calculations have uncovered analogous clathrate cages, but with inequivalent C:B ratios, identifying $I4/mmm$ CaB$_2$C$_4$ and SrB$_4$C$_2$ stoichiometry structures as superconductors with $T_c$s of 2 and 19~K, respectively, while SrB$_2$C$_4$ and BaB$_2$C$_4$ were predicted to possess superior mechanical properties with Vickers Hardnesses, $H_\text{v}$s, of 44 and 41~GPa, respectively~\cite{Cui:2022}, though they were not on the convex hull \cite{Li:2020sr}. We note that metal-doped clathrates with pure carbon frameworks are also predicted to exhibit high-$T_c$ superconductivity \cite{Zipoli:2006,Breda:2000c} (for example sodalite-like NaC$_6$ with a predicted $T_c$ above 100~K~\cite{Lu:2016}), however non-B-doped carbon clathrates have not been produced to date. We also note theoretical studies of the superconducting and mechanical properties of sodalite-like BN frameworks \cite{Li:2019, Hai:2022} and carbon-based materials~\cite{Lu:2016}, as well as metal hexaborides with a wide variety of measured properties including superconductivity\cite{Cahill:2019}.

In this article we employ the Density Functional Theory-Chemical Pressure (DFT-CP) method to study the effect of the metal atom's ionic size on the ambient-pressure dynamic stability of compounds in the $Pm\bar{3}n$ $X$B$_3$C$_3$ structure type ($X=$Mg, Ca, Sr, Ba). Moreover, we expand the metal atoms that can be stuffed within the clathrate framework to include the lanthanides. Turning to systems with two metal atoms on adjacent sites, we find 22 ordered alloy arrangements that are metastable at atmospheric pressures. Our theoretical study suggests that a wide range of possible elemental occupations and stoichiometries could be stabilized in this family of structures, potentially as solid solutions. The valence of the metal atom is shown to be a key factor in determining the superconducting critical temperature, $T_c$, which we predict could surpass the boiling point of liquid nitrogen at 1~atm, in particular for KPbB$_6$C$_6$. Finally, we illustrate that both configurational and vibrational entropy contributions, typically not considered in first principles calculations, are important in stabilizing these phases at pressure-temperature conditions accessible within diamond anvil cells.

\section{Methods}
Geometry optimizations of both the atomic positions and lattice parameters, molecular dynamics simulations, and electronic structure calculations including band structures, densities of states (DOS), and Bader charges were performed using density functional theory (DFT) as implemented in the Vienna \textit{ab-initio} Simulation Package (VASP) version 5.4.1 \cite{Kresse:1993a, Kresse:1999a}, with the gradient-corrected exchange and correlation functional of Perdew{-}Burke{-}Ernzerhof (PBE) \cite{Perdew:1996a}, and the projector augmented wave (PAW) method \cite{Blochl:1994a}. The plane-wave basis set energy cutoff was 600~eV for the compounds containing lanthanide atoms, and 900~eV otherwise. The $k$-point meshes were generated using the $\Gamma$-centered Monkhorst-Pack scheme and the number of divisions along each reciprocal lattice vector was selected so that the product of this number with the real lattice constant was 50~\AA{} in geometry optimizations and 70~\AA{} otherwise.
Phonon calculations were carried out on 2$\times$2$\times$2 supercells containing 112 atoms \cite{Parlinski:1997,Chaput:2011} using the VASP package coupled to the PHONOPY code\cite{Togo:2015}. The electron-phonon coupling (EPC) calculations were carried out using the Quantum Espresso (QE) program \cite{Giannozzi:2009}. The $T_c$s were estimated using the Allen-Dynes modified McMillan equation \cite{Allen:1975}, along with numerical solution of the Eliashberg equations \cite{Eliashberg:1960}. In the main text we discuss $T_c$s obtained for a renormalized Coulomb potential, $\mu^*=0.1$, and results computed for values of $\mu^*=0.13, 0.15$ and $0.17$ are provided in the Supplementary Information Tables S6-7.
A DFT-Chemical Pressure Analysis~\cite{Fredrickson:2012}, which visualizes the internal stresses inherent in a crystal structure as a consequence of steric constraints, was performed on select phases using the  \emph{CPpackage2}~\cite{Berns:2014a}.  Further computational details provided in Section S1. 

\section{Results and Discussion}
Density Functional Theory (DFT) calculations were carried out to investigate the borocarbide clathrates containing one or two electropositive elements, over a wide range of hole doping concentrations. Among the studied structures with a single element occupying the metal site (simple clathrates, with $X=$~Li, Na, K, Rb, Cs, Be, Mg, Ca, Sr, Ba, Sn, Pb, Sc, Y, La) only $Pm\bar{3}n$ $X$B$_3$C$_3$ ($X=$~Sr, Ba, Y, and La) phases were found to be dynamically stable at ambient pressure, in-line with the syntheses of the Sr and La analogues~\cite{Li:2020sr,Strobel:2021la}. Previous theoretical calculations concluded that Sr, Ba and La containing compounds were local minima at ambient pressure~\cite{Wang:2021, Di:2022}, and a mere 1~GPa was sufficient to stabilize CaB$_3$C$_3$~\cite{Di:2022}. The stability pattern of these simple clathrates illustrates the importance of a good size match between the borocarbide cage and the metal atom within. To further explore this relationship, DFT-Chemical Pressure (DFT-CP) schemes~\cite{Fredrickson:2012}, which reveal the internal tensions emerging in a crystalline lattice due to atomic size effects, were calculated for the alkaline earth metal borocarbide series. In DFT-CP calculations, comparisons between systems with component elements having similar pseudopotentials~\cite{Hilleke:2018} (in this case, metal atoms belonging to the same group of the periodic table) are most appropriate. This further allows us to isolate the role played by atomic size from other confounding factors, such as the oxidation state of the metal atoms. The $X$B$_3$C$_3$ alkaline earth metal borocarbides  (all nominally $X^{2+}$) offer a series for which, based on the emergence of dynamic stability when moving from Mg to Ba, atomic size is expected to play the main differentiating role. Recently, we have employed the DFT-CP method to elucidate the relationship between the metal atom size and the dynamic stability of related systems: $Pm\bar{3}n$  $X$B$_3$Si$_3$ ($X=$~Na, K, Rb, Cs)~\cite{Cui:2020} and $Im\bar{3}m$ $X$H$_6$~\cite{Zurek:2022f}. For those hydrides that did not correspond to local minima, DFT-CP was used to understand the structural distortions and chemical modifications that could be used to stabilize them~\cite{Zurek:2022f}. 

In the CP-schemes, shown for $X$B$_3$C$_3$ in Fig.~\ref{fig:MB3C3_CP}, negative pressures, which are indicative of an atom too small for its coordination environment, are represented with black, and positive pressures -- in white -- indicate an atom too large for the cavity within which it resides. For the alkaline earth borocarbides, the metal atom is in all cases surrounded by negative CP, suggesting that the borocarbide cages are plenty large enough -- and in some cases, too large -- to comfortably fit the electropositive atoms. This feature is most prominent within MgB$_3$C$_3$ (Fig.~\ref{fig:MB3C3_CP}(a)), which is dynamically unstable at ambient-pressure. The size of the CP lobe grows smaller as the size of the alkaline earth metal increases, hinting at a progressively better fit going down the group, in-line with predictions of dynamic stability at 1~GPa for CaB$_3$C$_3$ (Fig.~\ref{fig:MB3C3_CP}(b)) and at ambient-pressure for both SrB$_3$C$_3$ (Fig.~\ref{fig:MB3C3_CP}(c)) and BaB$_3$C$_3$ (Fig.~\ref{fig:MB3C3_CP}(d))~\cite{Wang:2021, Di:2022}.  This mirrors our findings for the $X$B$_3$Si$_3$ family, whose ambient-pressure dynamic stability was limited to the larger alkali metals Rb and Cs~\cite{Cui:2020}. 

The negative pressures on the metal atoms are balanced against positive pressures between the boron and carbon atoms, as the bipartite sodalite framework is drawn tight around them. The magnitudes of the positive pressures along the B-C contacts grow smaller when encapsulating larger metal atoms, reducing cramping in the borocarbide cage. Unlike in the superhydride analogues, the CP distributions around the boron and carbon atoms comprising the clathrate cage differ somewhat. The significant degree of electron transfer leads to the formation of anionic carbon and cationic boron held together by polar covalent bonds, increasing the effective sizes of the carbon atoms and decreasing those of boron. The differing CP distributions around boron and carbon atoms manifest as well in their vibrational properties, as we will soon see. Of course, this analysis accounts for only one factor contributing to dynamical stability in these compounds. As described above, none of the alkali metal borocarbides were dynamically stable at ambient pressure, in contrast to our previous findings on their borosilicide cousins RbB$_3$Si$_3$ and CsB$_3$Si$_3$. With typical oxidation states of +1, the alkali metal ions will be generally larger than the +2 alkaline earth metal ions. The dynamical stability of Rb and Cs within a borosilicide framework rather than a borocarbide one suggests a preference for the larger voids in the borosilicide network as C is replaced by Si. On the other hand, exploratory calculations on a series of borocarbides stuffed with lanthanide atoms -- with similar ionic radii -- showed that all considered were dynamically stable (Fig.\ S5-6). Factors such as the role of magnetism, spin-orbit coupling, and strong electron correlation are also expected to play a large role in the stability and electronic structure of these lanthanide borocarbides, and detailed calculations exploring these are underway.

Let us now turn to the borocarbides encapsulating two inequivalent metal atoms in adjacent cages (binary clathrates), with a focus on their stability patterns, electronic structure, and propensity for superconductivity.  Recently, Zhang and co-workers~\cite{Zhang:2022} carried out cluster expansion calculations coupled with CSP to determine the lowest enthalpy geometries resulting from the reaction of SrB$_3$C$_3$ and RbB$_3$C$_3$ to form Rb$_x$Sr$_{1-x}$B$_3$C$_3$ with different metal ratios ($x=0.125-0.5$), and considering different colorings (metal atom distributions~\cite{Burdett:1985}). Relevant for our study, they found that a $Pm\bar{3}$ RbSrB$_6$C$_6$ compound, wherein the shortest metal-metal contacts were between Rb and Sr, possessed the most negative enthalpies of formation (though for the CaYH$_{12}$ analogues, different colorings were preferred \cite{Zhao:2022p,Liang:2019}). Similarly, the most stable coloring for the Rb$_{0.4}$Sr$_{0.6}$B$_3$C$_3$ stoichiometry corresponded to a homogeneous distribution of the two metal atoms~\cite{Gai:2022}. With this in mind, we carried out high-throughput calculations on $Pm\bar{3}$ symmetry $XY$B$_6$C$_6$ borocarbides where the central metal atom in the structures shown in Fig.\ \ref{fig:MB3C3_CP} was of one type, and those on the corners of another type. In total 105 compounds were relaxed at ambient pressure, and their dynamic stabilities were evaluated via phonon calculations. Of the considered phases, 22 different combinations including $XY=$~KCa, KSr, RbSr, CaBa, SrBa, KPb, SrPb, BaPb, CaY, SrY, BaY, SnY, PbY, KLa, RbLa, CsLa, CaLa, SrLa, BaLa, SnLa, PbLa, and YLa were found to be dynamically stable. Molecular dynamic simulations in the $NPT$ ensemble suggested that all of these except for CaBa, SrBa, BaPb, BaY, CsLa, SnLa and YLa would be thermally stable at 300~K and 1~atm (Fig.\ S7-S11). Thus, it may be that these 7 phases are `fleeting' \cite{Hoffmann:2008} (decomposing quickly or persisting only at low temperatures), whereas the remaining 15 are likely to be closer to `viability' \cite{Hoffmann:2008} (with notable barriers to decomposition).

Zhang \emph{et al.} recently performed high-throughput calculations on 24 compounds in this structure type finding RbSr, KCa and KSr to be dynamically stable at ambient pressures~\cite{Zhang:2022}, and Di Cataldo and co-workers considered all 12 combinations of an alkali metal and an alkaline-earth metal, concluding that RbSr, KCa and NaSr were dynamically stable at ambient pressure, whereas CsBa became dynamically stable above 10~GPa~\cite{Di:2022}.  Therefore, in addition to the compounds identified as local minima by Zhang~\cite{Zhang:2022} and Di Cataldo (excluding NaSr)~\cite{Di:2022} and those that have been synthesized~\cite{Strobel:2021la,Li:2020sr}, we predicted a number of hitherto unreported dynamically stable phases involving Sn, Pb, Y, La as well as combinations of two alkaline earth metals at ambient pressure, expanding the number of possibly synthesizable $XY$B$_6$C$_6$ phases by 19 (18 metals, and 1 insulator), and investigated their thermal stability. 

As atomic and ionic radii are notoriously difficult to quantify in the setting of a compound, dependent as they are on specific local chemical environments, we identified the B-C distances in the borocarbide cages as easily extracted structural parameters to help rationalize our understanding of the dynamic stability of the $XY$B$_6$C$_6$ phases. These distances, informed by the ionic radii and valence of the metal atoms within, turned out to be important descriptors, with all structures corresponding to local minima having B-C interatomic distances that fell between 1.68-1.81~\AA{} (Fig.\ S12 and Tables S8-S11). In the $XY$B$_6$C$_6$ phases, the differing sizes of the $X$ and $Y$ atoms leads to two sets of B-C distances for each compound; for the dynamically stable phases the difference between these distances was smaller than 0.1~\AA{}. While some dynamically unstable systems have B-C distances adhering to these guidelines, we find the B-C distances nevertheless highlight the importance of a good size-match for the metal ions with one another, as well as for the borocarbide cages encapsulating them.

\begin{figure*}[ht!]
\begin{center}
\includegraphics[width=16cm]{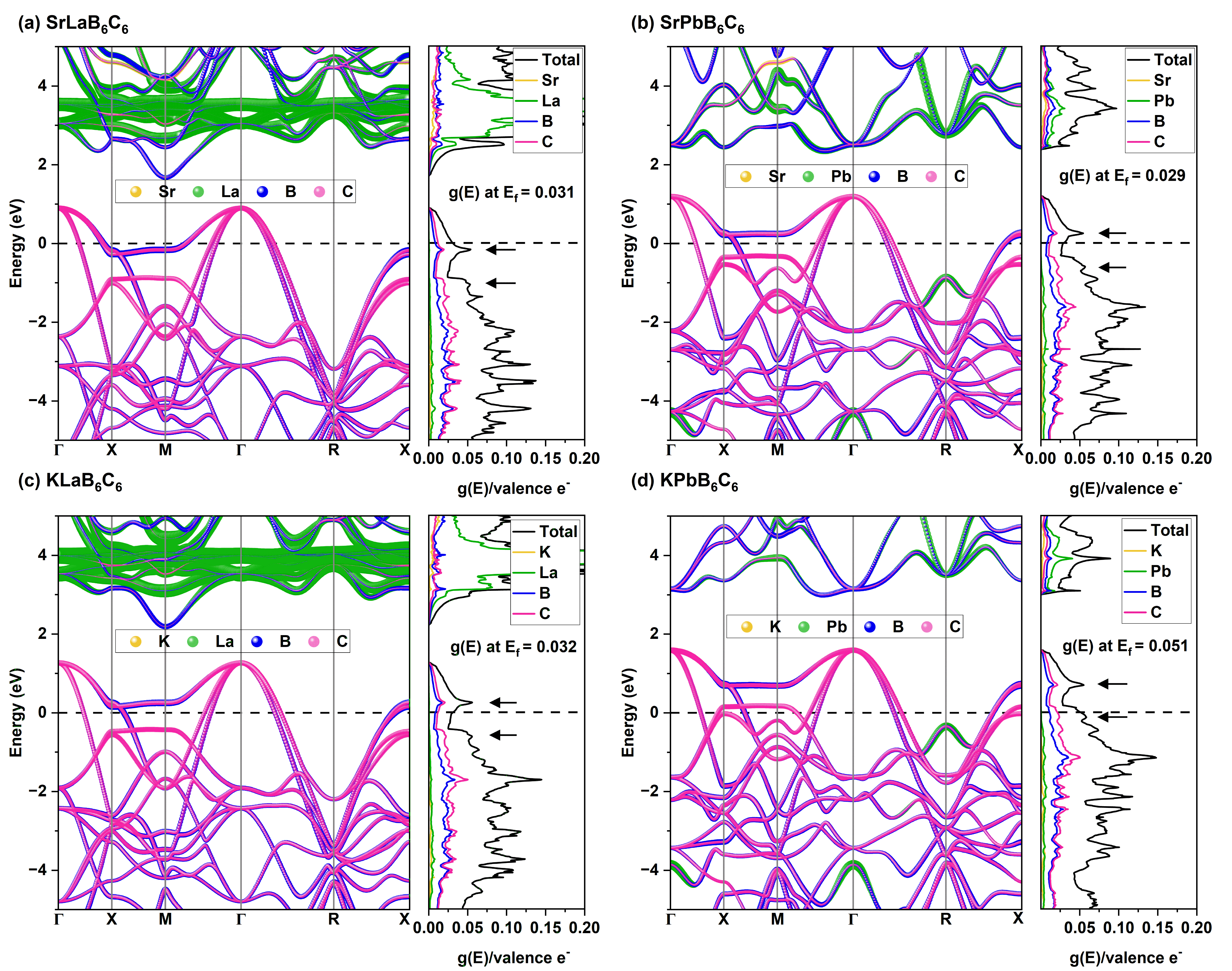}
\end{center}
\caption{Representative band structures and densities of states colored by their elemental characters for $XY$B$_6$C$_6$ phases containing two different metals with varying oxidation states. The $XY$ elements (average oxidation states of the metal atoms) are: (a) SrLa (+2.5), (b) SrPb (+2), (c) KLa (+2), (d) KPb  (+1.5). The Fermi energy is set to 0~eV, and arrows point to the peaks in the DOS plots. Pinning $E_F$ at these peaks would maximize the $T_c$.}  
\label{fig:DOS}
\end{figure*}

The intercalant modifies the number of states that can participate in the superconducting mechanism and the B-C distances, which in turn influence the vibrational properties and the electron-phonon coupling (EPC). Previous theoretical studies have shown that the superconductivity in this class of materials arises from motions associated with the clathrate framework, with the density of states (DOS) at the Fermi level ($E_F$) being a key descriptor for $T_c$~\cite{Di:2022,Gai:2022,Zhang:2022}. Assuming full electron transfer from the trivalent rare earths yields a formula of $X^{3+}$[B$_3$C$_3$]$^{3-}$,  where the borocarbide cage is isoelectronic to diamond, resulting in an insulator. Stuffing these borocarbides with electropositive elements whose average valence is less than +3 results in hole-doped metals. The band structure of this class of compounds exhibits rigid band behavior where the nature of the intercalant has a minimal effect on the shape of the bands, but can be used to precisely tune the placement of $E_F$. A maximal $T_c$ can be achieved when $E_F$ lies on one of the two peaks in the DOS plot below the gap in these electron-precise compounds (highlighted by arrows in Fig.\ \ref{fig:DOS})~\cite{Di:2022,Gai:2022,Zhang:2022}. Pinning $E_F$ at the upper peak in the DOS, as in Sr$_{0.75}$La$_{0.25}$B$_3$C$_3$, yields a calculated $T_c$ of 59~K~\cite{Gai:2022}, while pinning it at the lower peak, as in Rb$_{0.4}$Sr$_{0.6}$B$_3$C$_3$,  results in a $T_c$ of 83~K~\cite{Gai:2022}. The $T_c$s of ordered $XY$B$_6$C$_6$ alloys ($XY=$~RbSr, KCa and KSr~\cite{Zhang:2022,Di:2022}, and CsBa, NaSr~\cite{Di:2022}), where $E_F$ lies close to the lower peak in the DOS, were predicted to fall between 72-82~K at or near ambient pressures. Thus, the $T_c$ can be nearly doubled, for example, by replacing half of the alkaline earth atoms in SrB$_3$C$_3$ with Rb~\cite{Di:2022,Zhang:2022,Gai:2022}. 
 
We illustrate that the band structure is mostly invariant to the identity of the intercalant for four representative compounds containing two metal atoms of different valences in Fig.\ \ref{fig:DOS}.  In all of the borocarbides we considered, the states near $E_F$ were almost entirely due to the carbon-2p and boron-2p levels, with the former having a slightly larger contribution than the latter because of the enhanced electronegativity of carbon. The conduction bands, on the other hand, are mostly of boron-2p and metal character. Around 1.15 and 2.00~eV below the top of the valence band in electron precise YLaB$_6$C$_6$ (Fig.\ S13) lie two peaks that could be accessed by $\sim$0.75 and 1.55 hole doping, respectively.  The combination of a trivalent and a divalent metal atom (e.g.\ Sr$^{2+}$ and La$^{3+}$ with an average oxidation state of +2.5) moves $E_F$ down by about 1~eV, so it lies slightly above the first peak in the DOS (Fig.\ \ref{fig:DOS}(a)). While the most common oxidation states of lead are +2 and +4, our calculations consistently show this toxic element adopts the former valence in the binary clathrates. Thus, the average oxidation state on the metal atoms is +2 in SrPbB$_6$C$_6$ and in KLaB$_6$C$_6$, so in both compounds $E_F$ lies near the same position, falling within the pseudogap between the upper and lower DOS peaks (Fig.\ \ref{fig:DOS}(b,c)). Finally, KPbB$_6$C$_6$, with an average metal oxidation state of +1.5, possesses the highest DOS at $E_F$, which is pinned close to the top of the lower peak (Fig.\ \ref{fig:DOS}(d)). These maxima in the DOS result from the presence of two nearly parallel flat bands exclusively of boron and carbon character that are found along the $X-M-(\frac{1}{3})\Gamma$ high symmetry lines and are separated by an energy gap of 0.5-0.7~eV.  By choosing an appropriate combination of metal elements $E_F$ can be tuned to coincide with one of these two bands, thereby increasing the number of states that can participate in the superconducting mechanism and resulting in a ``flat-band / steep-band'' scenario~\cite{Simon:1997}, proposed to be key for superconductivity in MgB$_2$, and other compounds.~\cite{Bussmann:2016,Deng2005flat} The DOS at $E_F$ is about the same for $Pm\bar{3}$ $XY$B$_6$C$_6$ with $XY=$~SrLa, SrPb and KLa suggesting their predicted $T_c$s will fall close to that of SrB$_3$C$_3$, whereas for KPbB$_6$C$_6$ it is nearly double, tantalizing with the potential of a higher $T_c$.

\begin{figure*}[ht!]
\begin{center}
\includegraphics[width=15cm]{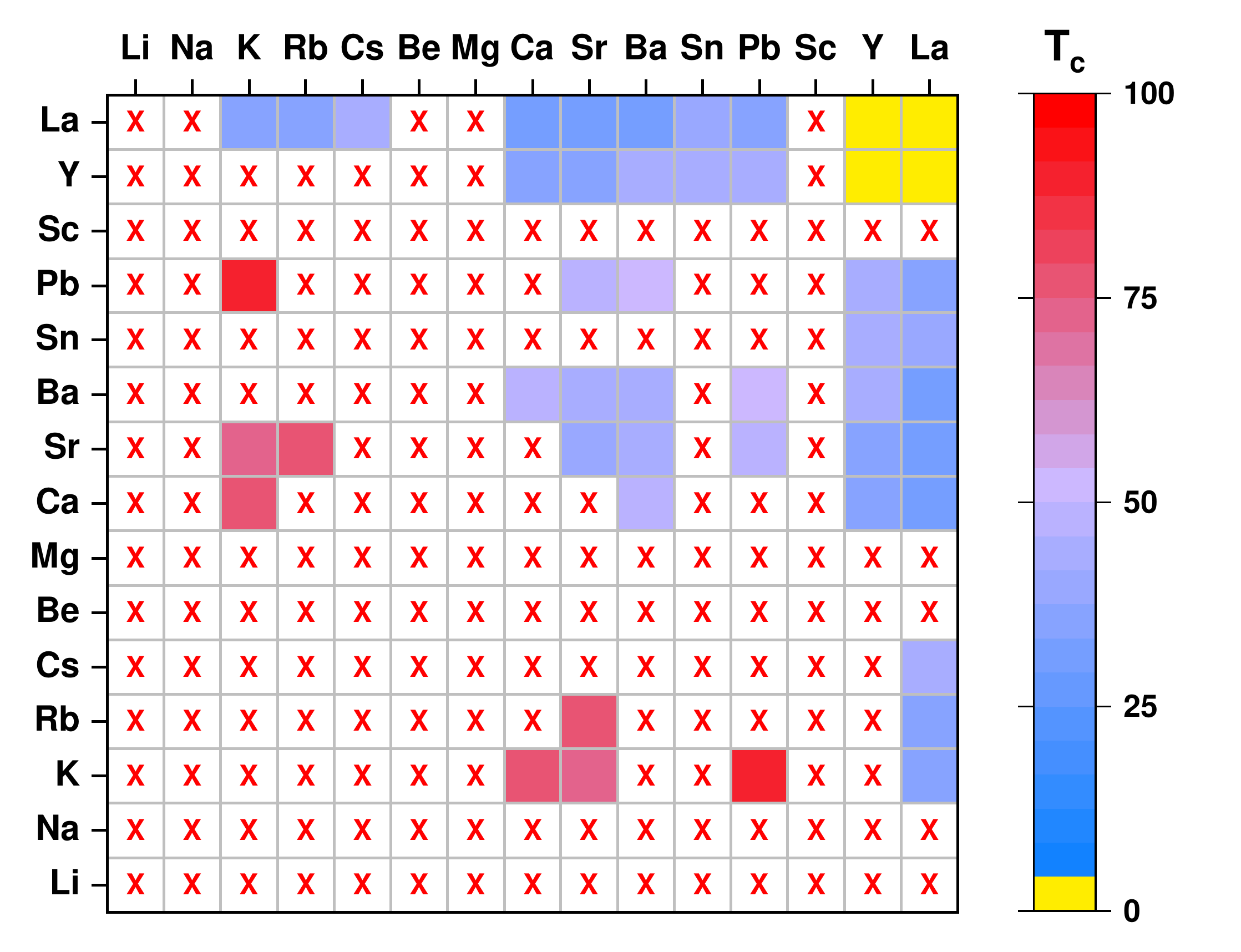}
\end{center}
\caption{A map of superconducting critical temperature, $T_c$, calculated via the Eliashberg equation for $Pm\bar{3}$ $XY$B$_6$C$_6$ phases at ambient pressure and assuming a Coulomb repulsion parameter, $\mu^*=$~0.1, where the identities of the metal atoms are given on the $x$- and $y$-axes. Phases that are dynamically unstable from phonon calculations are labelled with X's, and insulating phases are colored yellow.}
\label{fig:tc-map-MAE}
\end{figure*}

How do the identities of the metal atoms, their valences and the resulting DOS at $E_F$ influence the superconducting properties of the binary clathrates? To answer this question we calculated the Eliashberg spectral function,  $\alpha^2F(\omega)$, electron phonon coupling (EPC) constant, $\lambda$, and the logarithmic average frequency, $\omega_\text{log}$ (Tables S6-7), of the ambient-pressure dynamically stable compounds. The $T_c$s were estimated using the McMillan Allen-Dynes-modified (MAD) equation~\cite{Allen:1975} as well as via numerical solution of the (isotropic) Eliashberg equations~\cite{Eliashberg:1960}, which is more accurate for strongly coupled systems whose $\lambda \ge 1.5$. The Eliashberg results are summarized in map form in Figure \ref{fig:tc-map-MAE}, where dynamically unstable combinations are marked with a cross and dynamically stable phases are colored according to their $T_c$.  The plot illustrates results obtained for $\mu^*=0.1$, with those for $\mu^*=0.13, 0.15, 0.17$ provided in Table S7. The values we obtained for SrB$_3$C$_3$ (39~K) and BaB$_3$C$_3$ (45~K) are comparable to previous estimates (40-45~K\cite{Wang:2021, Di:2022,Zhang:2022,zhu:arxiv} and 43-50~K\cite{Wang:2021,Di:2022}, respectively).  Generally speaking, the phases with the lowest $T_c$s contained one divalent and one trivalent or one monovalent and one trivalent  metal atom ($T_c$~=~30-43~K), while superconductivity was predicted to persist to somewhat higher temperatures in borocarbide cages filled with divalent metal atoms ($T_c$~=~39-54~K), and combinations of a monovalent and a divalent atom remained superconducting until near the boiling point of liquid nitrogen ($T_c$~=~74-88~K),  perfectly correlating with the values of the DOS at $E_F$ calculated for the various metal combinations (Fig.\ \ref{fig:DOS}).

Delving into the trends more deeply, for SrB$_3$C$_3$ and BaB$_3$C$_3$ as well as most of the phases where the average oxidation state of the metal atoms was +2 ($XY=$~CaBa, SrBa, KLa, RbLa, and CsLa), $\lambda$ was found to lie between 0.86-1.12 and $\omega_\text{log}$ was between 501-588~K resulting in $T_c$s ranging from 36-48~K. The only exception consisted of two phases containing a Group 14 element, BaPb and SrPb, whose EPCs were 1.28 and 1.40, respectively. Though their $\omega_\text{log}$ was smaller, 400~K and 417~K, they still possessed a slightly higher $T_c$ (49~K for SrPb and 54~K for BaPb). 
The binary-metal borocarbides containing a divalent and a trivalent metal atom with an average oxidation state of +2.5 have not been studied before, however they represented the largest number of dynamically stable phases (Ca, Sr, Ba, Sn, and Pb in combination with La or Y). Their EPC ranged from 0.76-1.18 with the SnY and PbY combinations having the strongest coupling. Though some of these compounds possessed the highest $\omega_\text{log}$ of any of the borocarbides considered (CaLa, SrLa and BaLa with values of 625-653~K) this increase was overshadowed by the smaller $\lambda$ resulting in a $T_c$ that did not surpass the one calculated for BaB$_3$C$_3$. Once again, the $T_c$ was higher for the four phases that contained a Group 14 element (SnY, PbY, SnLa and PbLa).
In-line with their increased DOS at $E_F$ the EPC of the KCa, KSr and RbSr containing compounds, whose average metal oxidation state is +1.5, were among the largest ($\lambda=$~1.58, 1.61, 1.79). Their calculated Eliashberg $T_c$s (75, 74, 77~K) were similar to previous estimates of Zhang \emph{et.\ al.}~\cite{Zhang:2022} and Di~Cataldo and co-workers~\cite{Di:2022}. However, KPbB$_6$C$_6$ possessed the largest coupling of any of the systems studied so far, $\lambda=$~2.67, and concomitantly the highest $T_c$ of 88~K. Notably, the EPC in this phase was nearly the same as the value calculated for the first high pressure superhydride to be predicted, $Im\bar{3}m$ CaH$_6$, with $\lambda=$~2.69 at 150~GPa resulting in a predicted Eliashberg $T_c$ of 235~K~\cite{Ma:2012f}.

\begin{figure}[htb!]
\begin{center}
\includegraphics[width=17cm]{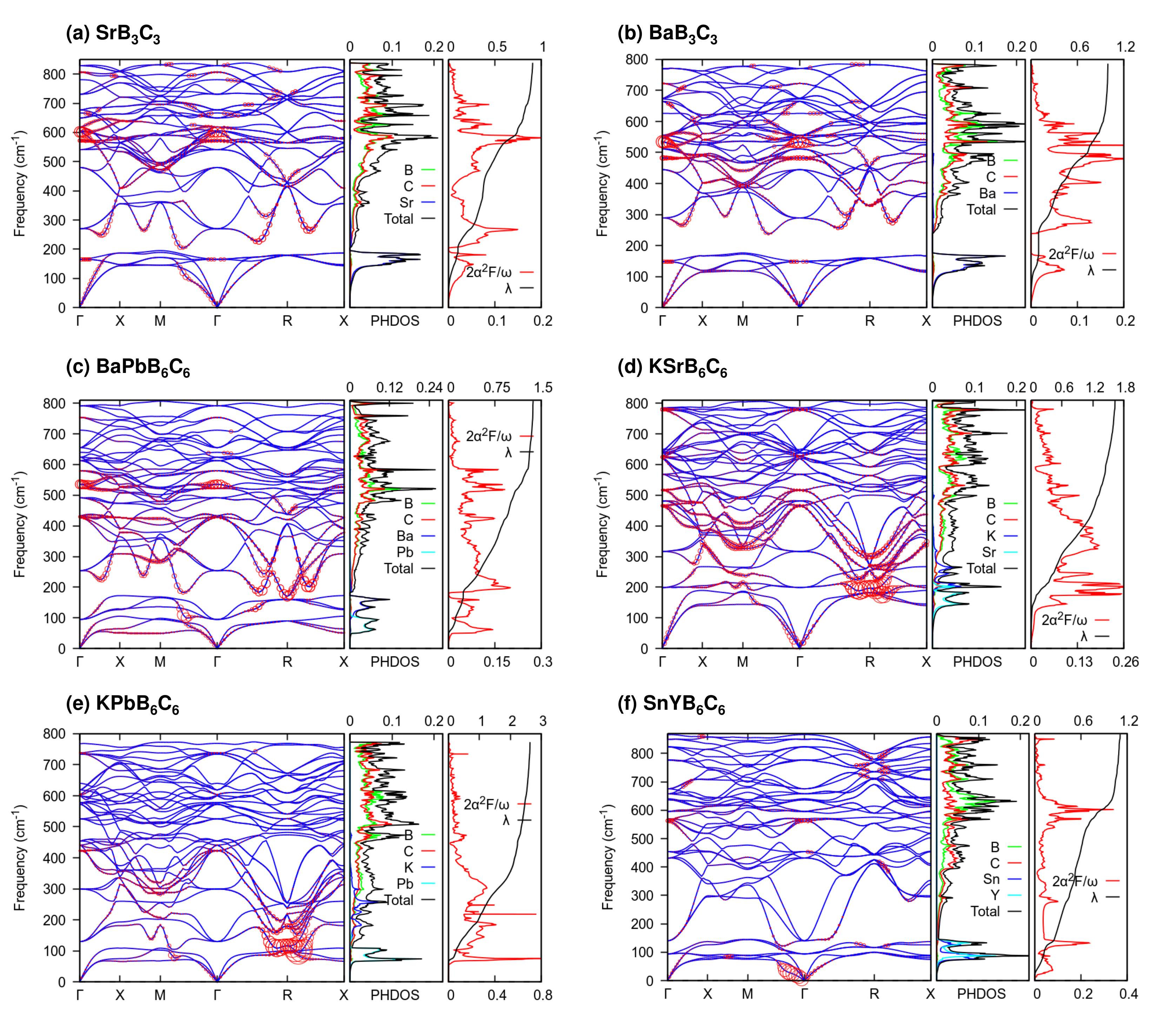}
\end{center}
	\caption{Phonon dispersion curves, projected phonon density of states (PHDOS), Eliashberg spectral function scaled by the frequency ($2\alpha^2F/\omega$), and the EPC integral ($\lambda(\omega)$) for a) SrB$_3$C$_3$, b) BaB$_3$C$_3$, c) BaPbB$_6$C$_6$, d) KSrB$_6$C$_6$, e) KPbB$_6$C$_6$, and f) SnYB$_6$C$_6$ at ambient pressure. The radius of the bubble on the phonon dispersion curve is proportional to the electron-phonon coupling constant ($\lambda_{\textbf{q}\nu}$) for the mode $\nu$ at wavevector \textbf{q}.} 
\label{fig:epc-spec}
\end{figure}

To analyze the phonon modes that are key for superconductivity, we plotted the phonon band structures, phonon (projected) densities of states, and Eliashberg spectral functions and EPC integrals, $\lambda(\omega)$, of all of the identified superconducting phases (Fig.\ S14-18). A subset of these are plotted in Fig.\ \ref{fig:epc-spec}: SrB$_3$C$_3$ and BaB$_3$C$_3$ (Fig.\ \ref{fig:epc-spec}(a,b)) were the simplest compounds considered, containing a single metal atom; $XY=$~BaPb (Fig.\ \ref{fig:epc-spec}(c)) possessed the highest $T_c$ for combinations of metals whose oxidation state averaged +2; KSr and KPb, with an average oxidation of +1.5, (Fig.\ \ref{fig:epc-spec}(d,e)) were characterized by $T_c$s approaching the boiling point of liquid nitrogen, and SnY (Fig.\ \ref{fig:epc-spec}(f)) had the highest $T_c$ for a +2.5 metal atom combination. For all of the phases considered the phonon spectra can be separated into two regions: the lower frequency modes (typically below 200~cm$^{-1}$ to 300~cm$^{-1}$) are mostly associated with the heavier electropositive atoms while the higher frequency modes can be attributed to boron and carbon atoms. The plots of $\lambda(\omega)$ showed that the higher frequency modes typically contributed to between $\sim$75-90\% of the total EPC (Table S6). One notable exception was the highest $T_c$ compound we found, KPbB$_6$C$_6$, where the contribution of the lower and higher frequency modes towards the overall EPC was about the same. This will be discussed in detail below.

In $Im\bar{3}m$ CaH$_6$ the EPC was derived primarily from the $T_{2g}$ and $E_g$ modes at the Zone center, which corresponded to the in-plane breathing and rocking vibrations of the H$_4$ square faces~\cite{Ma:2012f}. To determine which modes contributed to the EPC in the borocarbide analogues, red circles representing the size of the contribution to $\lambda$ at a particular wavevector $\textbf{q}$ and frequency $\nu$ were overlaid on the phonon band spectra (Fig.\ S14-18, Fig.\ \ref{fig:epc-spec}). For both SrB$_3$C$_3$ and BaB$_3$C$_3$ a mode at $\Gamma$ in which the boron atoms in the B$_2$C$_2$ square faces undergo a scissoring or B-C-B bending motion (601 and 534~cm$^{-1}$, respectively) exhibited the largest $\lambda_{\textbf{q}\nu}$ of any of the modes, while the analogous C-B-C bend (807 and 723~cm$^{-1}$) barely contributed to the EPC. For the isotypic, but dynamically unstable, MgB$_3$C$_3$ and CaB$_3$C$_3$ phases the scissoring mode of the carbon atoms comprising the square faces also possessed a higher frequency than the boron scissoring motion. The lower frequencies for the boron scissoring motions are reflected in the  larger negative CP lobes on the B atoms in Figure 1, compared to the smaller or absent negative CP lobes on the C atoms, as negative CP features are often aligned with softer phonon modes~\cite{Engelkemier:2016}. Wang and co-workers also identified these doubly degenerate $E_g$ modes, which couple with the C 2p states, as being the main contributors to the total $\lambda$ in SrB$_3$C$_3$ and BaB$_3$C$_3$~\cite{Wang:2021}. For most of the binary-metal borocarbides this mode also possessed a large coupling strength, though in some cases, most notably KPbB$_6$C$_6$, its contribution was negligible (Fig.\ \ref{fig:epc-spec}(e) and Table S12). 

In SrB$_3$C$_3$ Wang \emph{et al.}\ also identified a softening, or local decrease in frequency along an acoustic branch between $\Gamma$ and $M$ as being important in the EPC mechanism. In this mode the amplitude of the motion of the light elements was larger than that of the metal atoms despite the mode's low frequency~\cite{Wang:2021}. In our calculations on BaB$_3$C$_3$, however, this mode did not soften nor did it contribute to the EPC, in line with the results of Wang \emph{et al.} In binary clathrates, the set of low-lying optical phonons whose dispersion mimicked that of the acoustic modes could be associated with motions attributed to the lighter metal atoms. This is exemplified in Figure \ref{fig:epc-spec}c with the modes below 100~cm$^{-1}$ corresponding to Pb-based and the modes between 100-180~cm$^{-1}$ corresponding to Ba-based motions for BaPb. Certain combinations of metal atoms led to two sets of softened modes along the $\Gamma-M$ branch, one corresponding to each metal atom as in SrPb and KSr, while in others only one of these modes displayed a local softening (BaPb), and yet in others, such as RbSr, no local softening was observed in this portion of the phonon band structure. 

In addition to these previously discussed atomic motions that are important for the $T_c$, our calculations identified a number of modes that have not yet been scrutinized. For the simple clathrates this includes  the low-frequency optical mode, which softens around the midpoint of the $\Gamma-X$, $X-M$, $M-\Gamma$, $\Gamma-R$ and $R-X$ path and, for BaB$_3$C$_3$, a mode that softens substantially at the $R$-point where it is doubly degenerate. Curiously, at the $R$-point this vibration in BaB$_3$C$_3$ results in a Jahn-Teller like distortion of the near-perfect square faces into rectangular units with two short, and two long B-C distances. For many (but not all) of the binary clathrates with average +2 oxidation states, as well as some with an average oxidation state of +1.5, this mode is found to be highly softened at $R$ and a strong contributor to the EPC. A softening of the acoustic mode along some portion of the $X-M-\Gamma$ path was present in the KSr and KPb combinations, but not in RbSr or BaPb. Finally, for the four compounds with the highest EPC -- KSr, KPb, KCa, and RbSr -- softened acoustic modes along $\Gamma-R-X$, on either side of the $R$-point, and multiple soft modes at $R$ were found to be key for their very superb $T_c$.

The largest  $\lambda_{\textbf{q}\nu}$ in the highest $T_c$ compound we found, KPbB$_6$C$_6$, was a softened mode at 95~cm$^{-1}$ just off the $R$-point along the $\Gamma-X$ path, which could be described as a libration/wagging of the two boron and two carbon atoms on opposite corners of the square faces leading to the formation of a rectangle. Such motions are very similar to the aforementioned Jahn-Teller-like distortions. Above it, near 121~cm$^{-1}$ lay a mode with a slightly smaller $\lambda_{\textbf{q}\nu}$ whose main contributions could be attributed to similar libration and wagging motions in the B-C network, and along the $\Gamma-R$ path the light elements afforded the main contributions to a mode with intermediate $\lambda_{\textbf{q}\nu}$. These three modes -- at $R$, along $\Gamma-R$, and $R-X$ -- were all associated with coupled square-to-rectangle and square-to-tetrahedral motions of vertex-sharing four-membered faces on perpendicular planes. Thus, even though the low-frequency modes below 300~cm$^{-1}$ contributed to $\sim$48\% of the total $\lambda$, they could be primarily characterized by the motions of the boron and carbon atoms, mirroring the findings for SrB$_3$C$_3$~\cite{Wang:2021,zhu:arxiv}.

\begin{figure}[htb!]
\begin{center}
\includegraphics[width=17cm]{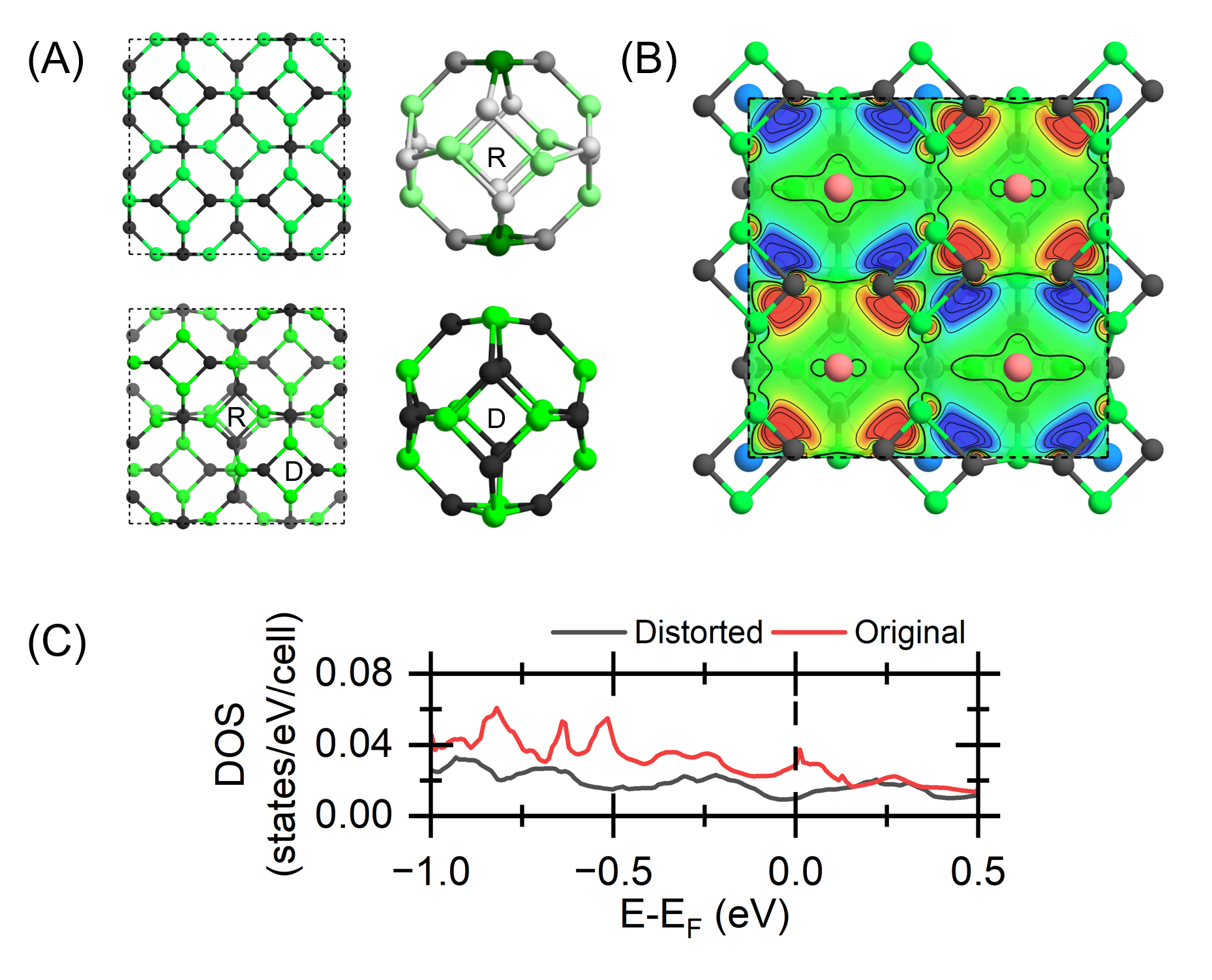}
\end{center}
	\caption{(a) A $2\times2\times2$ supercell showing the borocarbide framework of KPbB$_6$C$_6$ before (left panel, top) and after (left panel, bottom) a distortion by the $\sim$116~cm$^{-1}$ phonon mode at the $R$-point, which is associated with a significant EPC. The distortion yields three types of quadrilateral faces on the clathrate cage: square (not shown, perpendicular to the plane of the page), rectangular (R; right panel, top) and distorted-squares (D; right panel, bottom). (b) A contour plot obtained by subtracting the electron density of the distorted structure from that of $Pm\bar{3}$ symmetry  KPbB$_6$C$_6$ where red denotes a gain of charge (in the short bonds) and blue a loss of charge (in the long bonds). (c) The DOS near $E_F$ of the original and the distorted structure.}
\label{fig:displace}
\end{figure}

Let us further investigate the effect of the doubly degenerate soft mode at the $R$-point near 116~cm$^{-1}$ on the electronic structure of KPbB$_6$C$_6$. Closer examination of the structure resulting from displacing the atoms along the associated eigenvectors (Fig.\ S19 and Fig.\ 5(a) revealed it possessed some unperturbed square faces, others that were (nonplanar) distorted-squares, and some that became rectangular. A plot of the charge density difference  calculated by subtracting the electron density of the original structure from that of the perturbed geometry illustrated that no major charge transfer occurs along the B-C bonds comprising the square and distorted-square faces (Fig.\ \ref{fig:displace}(b)). For the B-C bonds along the rectangular faces, however, charge was transferred from the elongated to the shortened bonds, as expected, mirroring the change in the electronic structure that accompanies the Jahn-Teller distortion of C$_4$H$_4$ to cyclobutadiene. The distortion results in a decrease in the DOS at $E_F$ (Fig.\ \ref{fig:displace}(c)).

Given that SrB$_3$C$_3$~\cite{Li:2020sr} and LaB$_3$C$_3$~\cite{Strobel:2021la} have been synthesized under pressure we wondered if some of the $XY$B$_6$C$_6$ phases discussed above could be made using similar techniques and quenched metastably to ambient conditions? At zero pressure and temperature, the formation of ordered binary-metal borocarbides (from the elemental phases) is largely endothermic with computed $\Delta H_\text{f} > $150~meV/atom~\cite{Di:2022,Zhang:2022}. However, high pressures and temperatures could be used to access exergonic reaction pathways, for example the change in the free energy for the reaction $\text{Rb}+\text{SrB}_6+6\text{C} \rightarrow \text{RbSrB}_6\text{C}_6$ is negative by 3500~K  and 50~GPa~\cite{Zhang:2022}. Actually, the enthalpic instability of the binary-metal borocarbides could be expected: statistical analysis of the AFLOW data repository has shown that three or four component systems are typically only stabilized due to entropic factors that favor the formation of disordered systems \cite{Toher:2019}. One famous example is high entropy solid solutions, which are synthesized at high-temperatures but do not phase segregate when the temperature is decreased. Assuming a large (K,Pb)B$_3$C$_3$ supercell where there is no site preference for the two metal atoms, and using the Boltzmann formula (simplified via Stirling's approximation) we estimate the configurational entropy as $S/k_\text{B}=\ln\frac{N!}{(N/2)!(N/2)!}=\ln(1/2)$\cite{Rost2015}. This provides an upper bound to the stabilization that can be attained by randomly distributing the two metal atoms within the borocarbide cage. In a real system with some site preference the configurational entropy would be decreased; nonetheless we can use this value to estimate the conditions at which the configurational entropy would overcome the enthalpic penalty for the formation of (K,Pb)B$_3$C$_3$ from the elements. As shown in Table S13, we find this transition to occur by 4000~K and 50~GPa, suggesting that (K,Pb)B$_3$C$_3$ could potentially be made from the elements using laser heating within a diamond anvil cell (as could all of the metallic, dynamically stable binary clathrates we studied, see  Fig.\ S20-24 and Table S13), though other synthesis routes, not considered here, might also be viable. For example, because it is difficult to prepare potassium for high-pressure experiments, one could use binary compounds such as K$_2$C$_2$ or KB$_6$. Including the vibrational contributions to the free energy at 60~GPa within the harmonic approximation decreases the synthesis temperature further to 2000~K (Fig.\ S22(c)). 

Since $sp^3$ hybridized compounds containing boron and carbon, often in combination with a metal atom, are known to possess superior mechanical properties we used our machine learning approach~\cite{Zurek:2019b}, recently applied to predicted allotropes of boron and carbon~\cite{Zurek:2021b,Zurek:2022c}, to obtain the bulk and shear moduli, which, when combined with Teter's formula, can be used to estimate $H_\text{v}$ of the dynamically stable phases. As shown in Table S14, $H_\text{v}$ falls between 20-26~GPa, in-line with previous results for SrB$_3$C$_3$ based on calculated elastic properties~\cite{Li:2020sr} and comparable to measured values for tungsten carbide~\cite{Westbrook:1973}, and silicon carbide \cite{Milman:1999}. As expected, both the moduli and the Vickers hardnesses of the studied simple and binary borocarbides decrease with increasing metallicity. The $H_\text{v}$ of the insulating compounds (Table S14, Fig.\ S25) were among the  highest (24.6-28.5~GPa). Generally speaking, systems with an  average oxidation state of +2.5 possessed the second highest $H_\text{v}$s  (22.7-26.1~GPa), those with an average oxidation state of +2 were  predicted to be somewhat less hard ($H_\text{v}$=20.5-24.5~GPa) and the clathrates that had the highest DOS at $E_F$ with average oxidation states of +1.5  were the least likely to be hard ($H_\text{v}$=19.7-20.9~GPa). Thus, the DOS at $E_F$ in the simple or binary-guest borocarbides can be used to tune both the superconducting and mechanical properties of this family compounds, with an inverse correlation between the two.

\section{Conclusion}
In summary DFT-chemical pressure (DFT-CP) calculations have shown that the size of the electropositive alkaline earth metal atoms enclathrated within the bipartite sodalite borocarbide  cages,  $Pm\bar{3}n$ $X$B$_3$C$_3$, is a key factor for their ambient-pressure dynamic stability. The CP exerted by the larger metals on the cage within which they are confined elongates the B-C distances thereby minimizing the positive CP within the B-C network. Moving to borocarbide cages containing  two different metals with the $Pm\bar{3}$ $XY$B$_6$C$_6$ structure, from the 105 combinations considered, 22 ordered structures were found to be dynamically stable. From these, 19 are newly identified, 18 of which are new superconductors. The enclathrated metal atoms donate electrons to the borocarbide lattice and their average valence can be used to tune the number of occupied states at the Fermi level, $E_F$, so it falls on a peak in the density of states (DOS), thereby increasing the   ambient-pressure $T_c$ from 30~K in SrLaB$_6$C$_6$ to 88~K in KPbB$_6$C$_6$. Both the configurational and vibrational energy are shown to stabilize KPbB$_6$C$_6$, suggesting it  could potentially be synthesized at high temperatures above 50~GPa. Modes that are associated with distortions of the square faces in the borocarbide cages are important for the superconducting mechanism via electron-phonon coupling. Exploratory calculations on borocarbide cages stuffed with a single lanthanide element have uncovered a number of ambient-pressure metastable phases, suggesting that many more binary or even ternary metal combinations -- some whose $T_c$s may rival those of KPbB$_6$C$_6$ -- could potentially be made. The compounds predicted here, in addition to the large number of possible elemental occupations and stoichiometries not yet studied, provide a pathway towards warm superconductivity in covalent materials stabilized at ambient pressure.

\section*{Supporting Information}
The Supporting Information is available free of charge on the ACS Publication website. It includes the computational details, electronic band structures and densities of states, thermodynamic and thermal (molecular dynamics) stability analysis, Bader charges, structural parameters, Eliashberg spectral functions, phonon dispersion curves, EPC calculations, and details of the chemical pressure calculations. The structural parameters may also be obtained at Novel Materials Discovery\cite{nomad} (10.17172/NOMAD/2022.11.16-1) and The Cambridge Crystallographic Data Centre\cite{Parrinello:1980} (deposition number: 2220055 -- 2220080).

\section*{Acknowledgments}
We acknowledge the U.S.\ National Science Foundation for financial support, specifically grants DMR-2132491 (N.G.) and DMR-2119065 (X.W). K.H.\ is thankful to the U.S.\ Department of Energy, National Nuclear Security Administration, through the Capital-DOE Alliance Center under Cooperative Agreement DE-NA0003975 for financial support. Calculations were performed at the Center for Computational Research at SUNY Buffalo \cite{ccr}. Work at Carnegie was supported by the U.S.\ Department of Energy, Office of Science, Basic Energy Sciences, under Award Number DE-SC0020683. We acknowledge Warren Pickett for useful discussions.

\providecommand{\latin}[1]{#1}
\makeatletter
\providecommand{\doi}
  {\begingroup\let\do\@makeother\dospecials
  \catcode`\{=1 \catcode`\}=2 \doi@aux}
\providecommand{\doi@aux}[1]{\endgroup\texttt{#1}}
\makeatother
\providecommand*\mcitethebibliography{\thebibliography}
\csname @ifundefined\endcsname{endmcitethebibliography}
  {\let\endmcitethebibliography\endthebibliography}{}

\newpage

\textbf{Table of Contents Graphic}

\begin{figure*}
\begin{center}
\includegraphics[width=8.25cm]{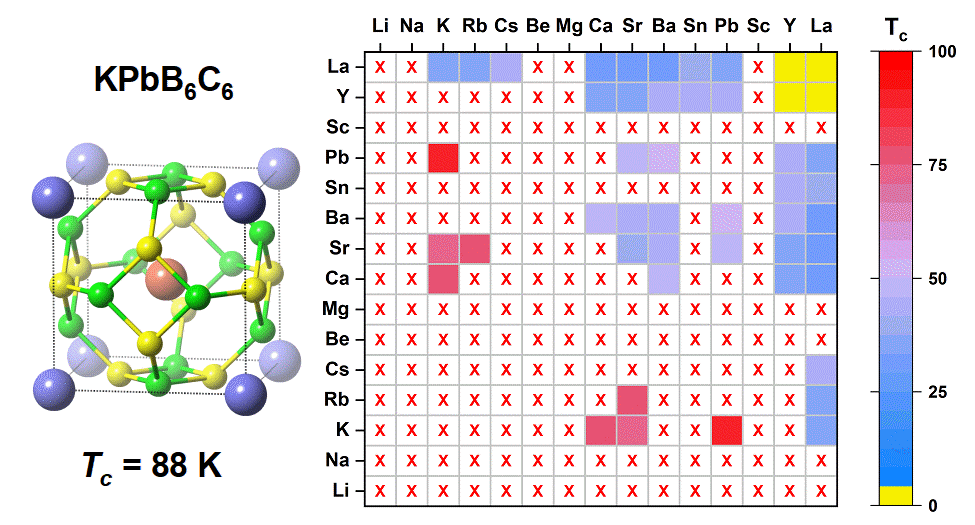}
\end{center}
\end{figure*}

\end{document}